\documentstyle[12pt,epsfig,twoside,fleqn,espcrc1]{article}

\newcommand{\AmS}{{\protect\the\textfont2
  A\kern-.1667em\lower.5ex\hbox{M}\kern-.125emS}}

\title{Intermediate mass dilepton production in heavy-ion collisions
at SPS energies\thanks{Work supported by the Department of 
Energy under grant No. DE-FG-88ER40388, by the Natural Science 
and Engineering Research Council of Canada, and by the Fonds 
FCAR of the Quebec Government.}}

\author{G.Q. Li\address{Department of Physics and Astronomy,  
        State University of New York\\
        at Stony Brook, Stony Brook, NY 11794, USA}
and C. Gale$^{\rm a}$\thanks{Permanent address: Physics 
Department, McGill University, Montreal, QC, H3A 2T8, Canada}}

\begin{document}

% typeset front matter
\maketitle

\begin{abstract}
Through the analysis of HELIOS-3 data on dilepton spectra, 
we demonstrate the importance of secondary processes 
for the dilepton production in heavy-ion collisions in the
intermediate mass region. We find that, while the dilepton 
spectra in proton-induced reactions 
can be nicely explained by the decay of primary vector mesons, 
charmed hadrons, and initial Drell-Yan processes, the strong 
enhancement seen in the heavy-ion data 
comes mainly from the secondary meson-meson interactions
which are unique to heavy-ion collisions. 
\end{abstract}

\vskip 0.5cm

The experimental measurement and theoretical investigation of
dilepton production in heavy-ion collisions constitutes one
of the most active and exciting fields in physics.
Because of their relatively weak final-state interactions
with the hadronic environment, dileptons are
considered ideal probes of the early stage of heavy-ion
collisions, where quark-gluon-plasma (QGP) formation and
chiral symmetry restoration are
expected.

Recent observation of the enhancement of low-mass dileptons in
central heavy-ion collisions by the CERES \cite{ceres,drees96} and
the HELIOS-3 \cite{helios} collaborations has generated a 
great deal of theoretical activities. The results from many 
groups with standard scenarios (i.e., using vacuum meson properties) 
are in remarkable agreement with each other, but in significant 
disagreement with the data: the experimental spectra in the 
mass region from 0.3-0.6 GeV are substantially underestimated 
\cite{drees96}. This has led to the suggestion of 
various medium effects that might be responsible for the
observed enhancement \cite{likob,rapp}. 

Another piece of interesting experimental data that has not received
much theoretical attention is the dilepton spectra in the
intermediate-mass region between 1 and 2.5 GeV.
Both the HELIOS-3 and NA38/NA50 collaborations have observed
a significant enhancement of dilepton yield in this mass region
in central S+W and S+U collisions as compared to that in the 
proton-induced reactions \cite{helios,na38}. 
The intermediate-mass dilepton spectra are particularly 
useful for the search of the QGP. It was originally suggested 
that in this mass region, the electromagnetic radiation 
from the QGP phase might shine over that from the hadronic 
phase. However, to extract from the 
measured dilepton spectra any information about the phase 
transition and the properties of the QGP, it is essential 
that the contributions from the hadronic phase be precisely 
understood and carefully subtracted.

In this contribution we report our recent study of dimuon spectra
in central S+W collisions \cite{ligale97}.
Previous thermal rate calculations 
show that in the mass and temperature region relevant for
this study, the following hadronic processes 
are important: $\pi\pi\rightarrow l{\bar l}$,
$\pi\rho\rightarrow l{\bar l}$, $\pi\omega\rightarrow l{\bar l}$,
$\pi a_1\rightarrow l{\bar l}$, $K{\bar K}\rightarrow l{\bar l}$, 
and $K{\bar K^*}+c.c \rightarrow l{\bar l}$ 
\cite{gale94,song94,haglin95,kim96}. The cross sections for 
the annihilation of pseudoscalar mesons are well known \cite{liko}.
The pion electromagentic form factor is dominated by the 
$\rho (770)$ meson, while that of the kaon is dominated by 
the $\phi (1020)$ meson. At large invariant masses, 
higher $\rho$-like resonances such as $\rho (1450)$ were 
found to be important. The cross sections for $\pi\rho\rightarrow 
l{\bar l}$ and ${\bar K}K^*+c.c.\rightarrow l{\bar l}$
have been studied in Ref. \cite{haglin95}.
High isoscalar vector mesons such as $\omega (1420)$ and $\phi (1680)$
play important roles in these processes. The cross section for 
$\pi \omega \rightarrow l{\bar l} $ has the same form as 
that for $\pi\rho\rightarrow l{\bar l}$, but with a different 
form factor.

The cross section for $\pi a_1\rightarrow l{\bar l}$ needs some 
special attention, since this process has been found to be
particularly important in the intermediate-mass region.
In Ref. \cite{gao97}, a comparative study was carried out 
for several existing models for the $\pi\rho a_1$ dynamics.
By using the experimentally-constrained spectral
function \cite{huang95}, it was found that the effective
chiral Lagrangian of Ref. \cite{comm84}
provides the best off-shell, as well as on-shell,
properties of the $\pi\rho a_1$ dynamics. The extraction of the form
factor for $\pi a_1\rightarrow l{\bar l}$ from $e^+e^-\rightarrow
\pi^+\pi^-\pi^+\pi^-$ still involves uncertainties in regarding
whether the $\rho (1700)$ resonance couples to $\pi a_1$ or not.
We will thus consider two scenarios for the 
$\pi a_1\rightarrow l{\bar l}$ form factor, one with and 
one without the $\rho (1700)$ contribution. 

For dilepton spectra with mass above 1 GeV, the contributions
from charm meson decay and the initial Drell-Yan processes
begin to play a role. These hard processes, however,
scale almost linearly with the participant nucleon number,
and can thus be extrapolated from the proton-proton and
proton-nucleus collisions. The results correponding to
the HELIOS-3 acceptance are shown in the left window of Fig. 1,
which are taken from Ref. \cite{drees96}. These, together
with the dileptons from the decays of primary vector mesons,
are collectively labeled `background'. It is seen that these
background sources describe very well the dimuon spectra in the
p+W reactions. However, as can be from the figure, the sum of 
these background sources grossly underestimates the dimuon 
yield in central S+W collisions, which indicates
additional sources to dilepton production in heavy-ion
collisions. This can come from the thermalized QGP and/or
hadronic phases. So the immediate next step is to check 
whether the contribution from the secondary hadronic processes
can explain this enhancement, which is the chief purpose of
this work. 

The contributions from the secondary processes outlined above
are shown in the middel panel of Fig. 1. These are 
obtained in the relativistic transport model of Refs. \cite{likob},
including the HELIOS-3 acceptances, mass resolution, and
normalization \cite{helios}. It is seen that the $\pi a_1$
process is by far the most important source for dimuon
yields in this mass region, as was found in many thermal 
rate calculations. The $\pi\omega$ process also play
some role in the entire intermediate mass region, while
the contributions from $\pi\pi$, $\pi\rho$ and $K{\bar K}$ 
are important around 1 GeV invariant mass.
In the right panel of Fig. 1, we compare the sum of the
secondary and background contributions
with the HELIOS-3 data for central S+W collisions. It
is seen that the data can now be nicely explained. Thus we
showed for the first time the importance of the secondary processes
for the intermediate-mass dilepton spectra in heavy-ion 
collisions. Although the current
data do not show any necessity to invoke the QGP formation
in S-induced reactions, consistent with conclusions 
from $J/\Psi$ physics, the observation that the secondary
processes do play an important role in the intermediate-mass
dilepton spectra is interesting and important. 

\begin{figure}[hbt]
\begin{center}
\vskip -1.0cm
\epsfig{file=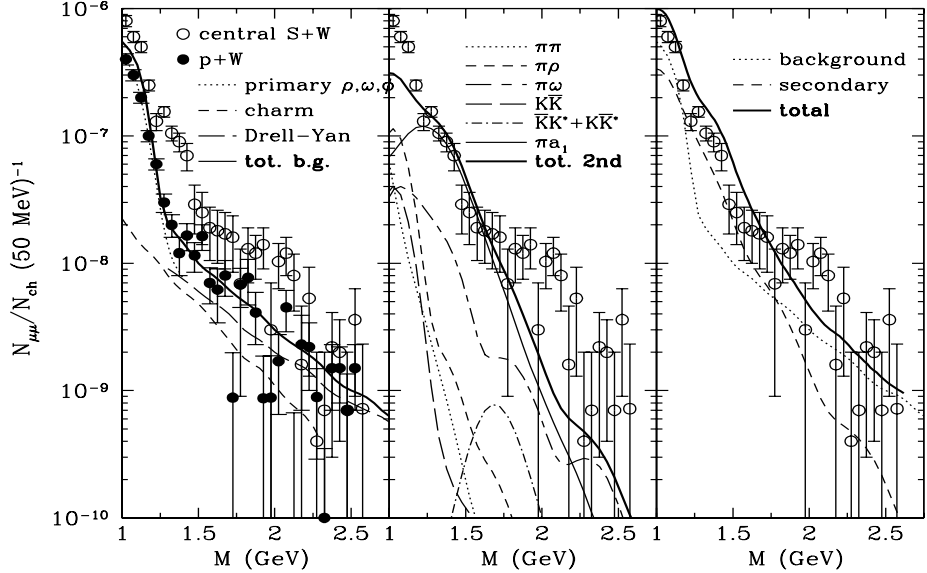,height=5.5in,width=3.5in,angle=270}
\vskip -0.6cm
\caption{Left panel: comparison of backgrounds with 
experimental data in p+W and S+W collisions.
Middle panel: contributions of various secondary
processes to the dimuom spectra in central S+W collisions.
Right panel: comparison of the sum of the background and 
secondary contributions with the experimental data in
central S+W collisions.}
\end{center}
\end{figure}

In the previous calculation we included $\rho (1700)$ in
the $\pi a_1$ form factor. we also did a calculation in which the 
$\pi a_1$ form factor contains only the normal $\rho (770)$.
From the formal point of view there is little evidence that
$\rho (1700)$ couples to $\pi a_1$.
The results are shown in the left panel of Fig. 2 by the dotted curve. 
Apparently, with a form factor that excludes the $\rho (1700)$
resonance, the contribution from the $\pi a_1$ process
is reduced. The agreement with the HELIOS-3 data in this
case is slightly better. Another issue we address 
is the effects of dropping meson masses 
on the dimuon spectra from the threshold to 
about 2.5 GeV. Below 1.1 GeV and especially from 0.4 
to 0.6 GeV, the agreement with the experimental data 
is much better when the dropping mass scenario 
is introduced. At higher masses, the
dropping mass scenarios somewhat underestimates the
experimental data. In this mass region, however, there might
be additional contributions from, e.g., secondary Drell-Yan 
processes \cite{spie97} that were not included in this study.

In summary, we have analysed the recent HELIOS-3 data 
on intermediate-mass dilepton production in heavy-ion 
collisions at the CERN SPS energies using the relativistic 
transport model. We have shown the importance of secondary processes 
for the dilepton production in heavy-ion collisions in this mass 
region. we found that $\pi a_1\rightarrow l{\bar l}$ to be the 
most important, as was found in many thermal rate calculations.

\begin{figure}[hbt]
\begin{center}
\vskip -1.0cm
\epsfig{file=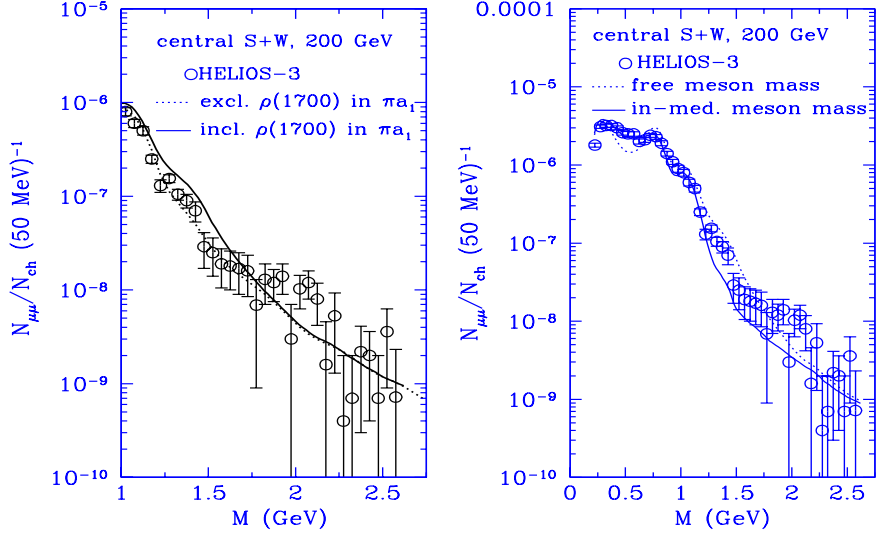,height=5.5in,width=3.5in,angle=270}
\vskip -0.8cm
\caption{Left panel: comparison of dimuon spectra obtained 
with and without the $\rho (1700)$ in the $\pi a_1$
form factor. Right panel: Comparison of dimuon spectra with
im-medium and vacuum meson masses.}
\end{center}
\end{figure}

The current investigation can be extended to higher incident energies,
such as those of RHIC collider, by combining the cross sections
(or thermal rates) obtained in this study with, e.g., 
hydrodynamical models for the evolution of heavy-ion collisions
at the RHIC energies. This kind study will be very useful for the
determination of hadronic background in the dilepton 
spectra, and for the clear identification of the dilepton
yield from the QGP.

\end{document}